\documentclass[jcp,twocolumn,amsmath,amssymb,floatfix,letter]{revtex4-1}
\usepackage{graphicx}
\usepackage{bm}

\newcommand{\ii}{{\rm i}}

\begin{document}

\title{Interatomic Methods for the Dispersion Energy Derived from the Adiabatic Connection Fluctuation-Dissipation Theorem}
\author{Alexandre Tkatchenko$^1$}
\author{Alberto Ambrosetti$^1$}
\author{Robert A. DiStasio Jr.$^2$}
\affiliation{$^1$Fritz-Haber-Institut der Max-Planck-Gesellschaft, Faradayweg 4-6, 14195, Berlin, Germany \\
$^2$Department of Chemistry, Princeton University, Princeton, NJ 08544, USA}

\begin{abstract}
Interatomic pairwise methods are currently among the most popular and accurate 
ways to include dispersion energy in density functional theory (DFT) calculations. However, 
when applied to more than two atoms, these methods are still frequently perceived to be
based on \textit{ad hoc} assumptions, rather than a rigorous derivation from quantum mechanics. 
Starting from the adiabatic connection fluctuation-dissipation (ACFD) theorem, an exact expression
for the electronic exchange-correlation energy, we demonstrate that the pairwise interatomic dispersion energy for 
an arbitrary collection of isotropic polarizable dipoles emerges from the second-order expansion of the ACFD formula. 
Moreover, for a system of quantum harmonic oscillators coupled through a dipole--dipole potential,
we prove the equivalence between the full interaction energy obtained from
the Hamiltonian diagonalization and the ACFD correlation energy in the random-phase approximation.
This property makes the Hamiltonian diagonalization an efficient method
for the calculation of the many-body dispersion energy.
In addition, we show that the switching function used to damp the dispersion interaction at short
distances arises from a short-range screened Coulomb potential, whose role
is to account for the spatial spread of the individual atomic dipole moments.
By using the ACFD formula we gain a deeper understanding of the approximations made in the interatomic pairwise
approaches, providing a powerful formalism for further development of accurate
and efficient methods for the calculation of the dispersion energy.
\end{abstract}

\maketitle

\section{Introduction}
Van der Waals (vdW) forces are ubiquitous in nature, and they play a major role in 
determining the structure, stability, and function for a wide variety of systems, including
proteins, nanostructured materials, as well as molecular solids and liquids. 
A significant attractive part of the vdW energy corresponds to the dispersion energy,
which arises from correlated fluctuations between electrons.
Therefore, accurate treatment of the dispersion energy is essential for improving our understanding 
of biological and chemical systems, as well as (hard and soft) condensed matter systems in general.
Many encouraging ideas and methods have been proposed in recent years 
for approximately including the missing long-range dispersion interactions in density functional
theory (DFT)~\cite{Hobza-review,Kannemann-Becke,Grimme-review,LL-review,Corminboeuf-2012,MRS-Bull}.
Despite significant progress in the field of modeling vdW interactions during the 
last decade, many questions still remain unanswered and further development is required 
before a truly universally applicable (accurate and efficient) method emerges. 
For example, interatomic vdW potentials are frequently employed for the modeling
of hybrid inorganic/organic interfaces~\cite{McNellis-PRL,Bluegel-PRL,Tonigold-Gross,graph-metal-PRL},
neglecting the relatively strong long-range Coulomb screening present within inorganic bulk materials.
On the other hand, the popular non-local vdW-DF functionals~\cite{vdW-DF04,vdW-DF10,VV10-review} use a homogeneous
dielectric approximation for the polarizability, which is not expected to be accurate for molecules.
Nevertheless, the interaction energies between small organic molecules turn out to be reasonably accurate.
Understanding the physical reasons of why these different approaches ``work well''
outside of their expected domain of applicability is important for the 
development of more robust approximations. 

Interatomic pairwise dispersion approaches, typically dubbed DFT-D~\cite{Grimme-review} or DFT+vdW~\cite{TS-vdw}, are among the
most widely used methods for including dispersion energy in DFT.
Such approaches approximate the dispersion energy in a pairwise fashion, \textit{i.e.} as a sum over unique atom pairs.
Despite their simplicity, these effective pairwise models provide remarkable accuracy when applied to small molecular systems,
in particular when accurate dispersion coefficients ($C_n$) are employed for atoms in molecules~\cite{jctc-2011}. 
Only recently have efforts been focused on going beyond the effective pairwise treatment of vdW contributions, for example,
the role of the non-additive three-body interatomic Axilrod-Teller-Muto term was assessed~\cite{Grimme-D3,me-Anatole-2010}.
Furthermore, an efficient and accurate interatomic many-body dispersion (MBD) approach to dispersion 
interactions has recently been proposed~\cite{MBD}. 
The MBD description of vdW interactions is essential for the description of extended molecules and molecular solids;
however, the influence of MBD interactions can already become significant when considering the interactions between relatively small 
organic molecules~\cite{MBD,PNAS-2012}.

Despite the popularity and the relative accuracy of the DFT-D and DFT+vdW methods, 
they are still widely perceived to be based on \textit{ad hoc} assumptions.
For the dispersion interaction between two spherical atoms $i$ and $j$, the pairwise $C_6^{ij} R_{ij}^{-6}$ 
formula has been known since the seminal work of London~\cite{London,Margenau-RMP}. However, to the best of our knowledge, 
the generalization of London's formula for an arbitrary collection of $N$ spherical atoms 
has not been explicitly derived from first principles.
Furthermore, at short interatomic distances, the dispersion interaction is significantly weaker than the
corresponding asymptotic expansion, and \textit{ad hoc} approximations become necessary for 
the functional form of the damping.
It is shown in this work that the switching function used to damp the dispersion interaction at short
distances arises from a short-range screened Coulomb potential, whose physical role
is to account for the spatial spread of the individual atomic polarizabilities.

Since the dispersion energy arises from correlated fluctuations between electrons, it is intrinsically a many-body phenomenon, an accurate 
description of which requires a quantum mechanical treatment. 
For this purpose, the adiabatic connection fluctuation-dissipation (ACFD) theorem, which provides a general and exact expression for the 
exchange-correlation energy~\cite{ACFDT1,ACFDT2}, allows us to calculate the dispersion energy in a seamless and accurate fashion, naturally 
including many-body effects.
In this work, we show that the ACFD theorem provides a firm theoretical basis for the development and understanding of interatomic pairwise and many-body 
dispersion methods. In particular, we derive the well-known $C_6/R^6$ interatomic pairwise summation formula as the second-order expansion of the ACFD correlation 
energy, and demonstrate that this formula is valid for an arbitrary collection of $N$ fluctuating dipoles, each of which 
is characterized by an individual frequency-dependent polarizability.
By applying the ACFD formalism we also prove, for a system of quantum harmonic oscillators (QHOs) coupled within the dipole approximation,
the mathematical equivalence between the exact dispersion energy and the correlation energy in the random-phase approximation (RPA). 
This analytical result makes the coupled-oscillator model (with a relatively minimal computational cost) an ideal candidate for the inclusion of MBD effects in DFT.
Finally, we show the relevance of MBD energy on the binding energies of dimers in the S22 database. The 
full many-body description consistently reduces mean relative errors with respect to the interatomic pairwise approximation, 
showing the largest improvements for the most extended systems.
The ACFD formula leads to a deeper understanding of the approximations made in the 
development of the DFT-D and DFT+vdW approaches, and provides a 
natural formalism for further improvement of methods
for computing the dispersion energy.

\section{The pairwise interatomic dispersion energy from the ACFD formula}  
The ACFD theorem is an exact expression for the exchange-correlation energy
of a system of nuclei and electrons, described by a response function
$\chi({\bf r},{\bf r'},\omega)$~\cite{ACFDT1,ACFDT2}. The response function $\chi$ measures
the response at point ${\bf r}$ due to a change of the potential at point ${\bf r'}$ as
a function of time or (Fourier-transformed) frequency $\omega$.
Here our interest lies in the dispersion energy, which is contained
in the electron correlation energy. Therefore we start by writing
the ACFD formula exclusively for the correlation energy
(Hartree atomic units are used throughout)
\begin{equation}
\label{eqACFD}
E_c = -\frac{1}{2\pi} \int_{0}^{\infty}d\omega \int_0^{1}d\lambda {\rm Tr}[(\chi_{\lambda}-\chi_0)v] ,
\end{equation}
in which $\chi_0$ is the bare or non-interacting response function, which can be computed given a set of single-particle
orbitals~\cite{Adler:1962,Wiser:1963} $\phi_i$ (with corresponding eigenvalues $\epsilon_i$ and occupation numbers $f_i$) as
\begin{equation}
\label{eqchi0}
\chi_0({\bf r},{\bf r'},\omega)=2 \sum_{i,j}(f_i-f_j) \frac{\phi_i^*({\bf r})\phi_i({\bf r}')\phi_j^*({\bf r}')\phi_j({\bf r})}{\epsilon_i-\epsilon_j+i\omega} ,
\end{equation}
$\chi_{\lambda}$ is the interacting response function at Coulomb coupling strength $\lambda$, defined \textit{via} the self-consistent Dyson
screening equation $\chi_{\lambda} = \chi_0 + \chi_0 (\lambda v + f_{xc}^{\lambda}) \chi_{\lambda}$,
$v = |{\bf r} - {\bf r'}|^{-1}$ is the Coulomb potential, and $\rm{Tr}$ denotes the trace operator
over spatial variables ${\bf r}$ and ${\bf r'}$. Using the ACFD formula, the adiabatic connection between 
the non-interacting system (with $\lambda=0$) and the fully interacting system (with $\lambda=1$),
yields the full correlation energy of the system of interest. Obviously, the correlation energy obtained
from the ACFD formula contains the full many-body dispersion energy as well as other electron correlation effects. 

In practice, the exact form of the exchange-correlation kernel $f_{xc}^{\lambda}$ in the Dyson equation is not known. Neglecting the explicit dependence
of $f_{xc}$ on $\lambda$, analytic integration can be carried out over $\lambda$. 
This is the case, for example, for the widely employed random-phase approximation (RPA)~\cite{Bohm-Pines,GellMann-Brueckner},
or the full potential approximation (FPA)~\cite{LangLunq}. In the RPA, $f_{xc}=0$, while in the FPA
$\chi_{\lambda} = \chi_1$, \textit{i.e.} the $\lambda$ integration is carried out using the fully interacting
response function. In what follows, we will employ the RPA method, which has been shown to yield reliable results for 
a wide variety of molecules and extended systems~\cite{Fuchs/Gonze:2002,Furche-vanVoorhis,
Furche-RPA,Scuseria/Henderson/Sorensen:2008,Janesko/Henderson/Scuseria:2009,Toulouse/etal:2009,Harl/Kresse:2008,
Harl/Kresse:2009,Lu/Li/Rocca/Galli:2009,Dobson/Wang:1999,Rohlfing/Bredow:2008,Ren/Rinke/Scheffler:2009,Schimka/etal:2010,
Galli-JPCA-2010,Ren/etal:2011}.
Using the Dyson equation, the ACFD correlation energy expression in Eq.~(\ref{eqACFD}) takes on the following form in the RPA
\begin{equation}
\label{eqRPAl}
E_{c,\rm{RPA}} = -\frac{1}{2\pi} \int_{0}^{\infty}d\omega \int_0^{1}\frac{d\lambda}{\lambda} {\rm Tr} \left[ \frac{(\lambda v \chi_0)^2}{1-\lambda v \chi_0} \right].
\end{equation}
Integration over $\lambda$ in Eq.~(\ref{eqRPAl}) leads to the following
expansion for the correlation energy in terms of $\chi_0 v$~\cite{Galli-JCP-RPA-expansion,Xinguo-beyond-DFT}
\begin{equation}
\label{eqRPA}
E_{c,\rm{RPA}} = - \frac{1}{2\pi} \int_{0}^{\infty}d\omega \sum_{n=2}^{\infty} \frac{1}{n} {\rm Tr}[(\chi_0 v)^n] .
\end{equation}

Let us now apply the ACFD-RPA approach to a collection of fluctuating dipoles representing the atomic system of interest. 
Each atom $i$ is characterized by its position $r_i = \{x_i, y_i, z_i\}$ and a frequency-dependent dipole polarizability $\alpha_i(\ii\omega)$.
For the moment, we assume that the atoms are separated by a sufficiently large distance, allowing us to use the bare 
Coulomb potential to describe the interaction between dipoles. The general case will be addressed in the next section.
The response function for each atom $i$ takes the form~\cite{Dobson-book}
\begin{equation}
\label{eqChiPol}
\chi_i({\bf r},{\bf r'},\ii\omega) = - \alpha_i(\ii\omega) \nabla_{\mathbf{r}}\delta^3({\bf r} - {\bf r_i}) \otimes \nabla_{\mathbf{r'}}\delta^3({\bf r'} - {\bf r_i}) ,  
\end{equation}
where $\delta$ is the three-dimensional Dirac delta function, and $\otimes$ is the tensor (outer) product.  

Now it will be shown that the second-order ($n=2$) term in Eq.~(\ref{eqRPA})
yields the pairwise interatomic dispersion energy. 
In the remainder of this section we drop the $\lambda$
index of the response function, because the atomic polarizabilities can be derived from a mean-field or
an explicit many-body calculation. Furthermore, the second-order $n=2$ term in Eq.~(\ref{eqRPA})
turns out to be the same if the expansion is carried out in terms of $\chi_0$ or $\chi_1$~\cite{Galli-JCP-RPA-expansion}. 
For a collection of $N$ atoms in the dipole approximation, the $\chi v$ matrix can be written as $AT$. Here, $A$ is
a diagonal $3N\times3N$ matrix, with $-\alpha_i(\ii\omega)$ values on the diagonal blocks. The $T$ matrix
is the dipole-dipole interaction matrix, with 3$\times$3 $ij$ tensors given by 
$T_{ij} = \nabla_{{\bf r_i}} \otimes \nabla_{{\bf r_j}} v_{ij}$ ($T_{ii} = 0$). 
For two atoms ($\alpha_1(\ii\omega)$ and $\alpha_2(\ii\omega)$) separated by a distance $R=|{\bf r_1} - {\bf r_2}|$ on the $x$ axis, the $AT$ 
matrix becomes
\begin{equation}
\label{atmatrix}
\left( \begin{array}{cccccc}
0 & 0 & 0 & \frac{2\alpha_1(\ii\omega)}{R^3} & 0 & 0 \\
0 & 0 & 0 & 0 & -\frac{\alpha_1(\ii\omega)}{R^3} & 0 \\
0 & 0 & 0 & 0 & 0 & -\frac{\alpha_1(\ii\omega)}{R^3} \\
\frac{2\alpha_2(\ii\omega)}{R^3} & 0 & 0 & 0 & 0 & 0 \\
0 & -\frac{\alpha_2(\ii\omega)}{R^3} & 0 & 0 & 0 & 0 \\
0 & 0 & -\frac{\alpha_2(\ii\omega)}{R^3} & 0 & 0 & 0 
\end{array} \right) .
\end{equation}
The second-order ($n=2$) term of the ACFD-RPA expression in Eq.~(\ref{eqRPA}) with the above matrix as 
input leads to 
\begin{equation}
\label{eqC2atoms}
E_{c,\rm{RPA}}^{(2)} = -\frac{1}{2\pi} \int_0^{\infty} d\omega \alpha_1(\ii\omega) \alpha_2(\ii\omega) {\rm Tr}[{T_{12}}^2] = - \frac{C_6^{12}}{R^6} ,   
\end{equation}
where the Casimir-Polder identity has been used to determine $C_6^{12}$ from the corresponding dipole
polarizabilities. The above equation is the familiar expression for the long-range dispersion 
interaction between two atoms. 

Equation~\ref{eqC2atoms} can be easily generalized to the case of $N$ atoms.
The use of the trace operator in Eq.~(\ref{eqRPA}) requires multiplication of column $i$ of 
the $AT$ matrix by the corresponding row $i$. 
Since for any given $T_{ij}$, ${\rm Tr}[{T_{ij}}^2] = 6/R_{ij}^6$, where
$R_{ij}$ is the distance between atoms $i$ and $j$, the second-order expansion of Eq.~(\ref{eqRPA}) for 
$N$ atoms leads to this simplified form 
\begin{equation}
\label{eq2Body}
E_{c,\rm{RPA}}^{(2)} = - \frac{1}{2} \sum_i \sum_j \frac{C_6^{ij}}{R_{ij}^6}.
\end{equation} 
This of course is the familiar expression for the dispersion energy for an assembly of $N$ atoms used in the DFT-D and DFT+vdW methods.   
We note that the above derivation of the pairwise dispersion energy from the ACFD formula does not make 
any assumptions regarding the geometry of the atomic assembly or the functional form of the frequency-dependent 
polarizabilities.
Furthermore, we note that employing the FPA instead of the RPA would not change the conclusions presented above.

While the second-order expansion of Eq.~(\ref{eqRPA}) for isotropic polarizabilities yields the 
familiar pairwise interatomic dispersion energy given by Eq.~(\ref{eq2Body}), the former equation is more general. 
It allows for the use of full polarizability tensors, enabling an anisotropic treatment of the dispersion
energy. In this regard, the polarizability anisotropy has been found to play a non-negligible 
role for intermolecular dispersion interactions~\cite{Geerlings-anisotropy,Buckyball-catcher}.

We note that the higher-order terms in the RPA expansion of the correlation energy include two 
contributions: higher-than-pairwise many-body interactions (up to $N$-th order) and the response screening (up to \textit{infinite} order). 
As an example of the beyond-pairwise many-body interactions captured in the RPA expansion of the correlation energy, the third-order term includes 
the well-known Axilrod-Teller-Muto three-body energy~\cite{AxilrodTeller}. The higher-order response screening can be easily illustrated for two interacting 
atoms $i$ and $j$ by expanding Eq.~(\ref{eqRPA}) (explicit dependence of the polarizabilities on $\ii\omega$ assumed):
\begin{equation}
\label{eqinfresp}
E_{c,\rm{RPA}} = -\frac{1}{2\pi} \int_0^{\infty} d\omega \left ( \frac{6\alpha_i\alpha_j}{R_{ij}^6} + 
\frac{36\alpha_i^2\alpha_j^2}{R_{ij}^{12}} + \ldots \right ) ,
\end{equation} 
in which the second-order term corresponds to the ``standard'' $C_6/R^6$ dispersion interaction between $i$ and $j$, and the higher-order terms
(which only survive with even powers of $n$) correspond to the response screening of the polarizability of atom $i$ by the presence
of atom $j$ and vice versa. Further analysis of these many-body contributions is presented in Sections IV and V.


\section{The damping of the dispersion energy at short distances}
Correlation energy calculations carried out using the ACFD formula usually
employ the response function $\chi_0$ computed using a set of occupied and virtual one-particle orbitals [see Eq.~(\ref{eqchi0})],
determined from semilocal DFT, Hartree-Fock, or hybrid self-consistent-field calculations. 
In this scenario, $\chi_0$ is typically a fairly delocalized object, which includes
orbital overlap effects between occupied and virtual states. However,
even in this case, the use of certain approximations for the exchange-correlation
kernel $f_{xc}$ can lead to divergencies for close inter-particle separations~\cite{Furche-vanVoorhis}. 
When the response function is localized, leading to a diagonal form, the details
of the overlap between orbitals are lost. For example, this is clearly the case
for an assembly of fluctuating point dipoles. When two point dipoles come into close contact,
the Coulomb interaction between them diverges. In fact, depending on the absolute values of the polarizability, 
the head-to-tail alignment between two dipoles can lead to an infinite polarizability even for a \emph{finite} 
(non-zero) separation between the dipoles~\cite{Stone-book}. This is clearly an unphysical situation,
which is mitigated in practice by the finite extent of electronic orbitals.
From a slightly different perspective, the dipole moment should be spread out in space, 
and this effect naturally introduces damping when the polarizabilities overlap. 

The most straightforward way to incorporate overlap effects for a set of fluctuating
point dipoles is through a modification of the interaction potential.
Thus, instead of using the bare Coulomb potential $v = |{\bf r} - {\bf r'}|^{-1}$,
a modified potential should be employed that accounts for orbital overlap at short distances. 
We take the polarizability of each atom $i$ to correspond to a quantum harmonic oscillator (QHO)
\begin{equation}
\label{eqQHO}
\alpha(\ii\omega) = \frac{\alpha_0}{1 + (\omega/\omega_0)^2} ,
\end{equation} 
where $\alpha_0$ is the static dipole polarizability, and $\omega_0$ is
an effective excitation frequency. Since the ground state QHO wavefunction has a gaussian form, the interaction between
two QHOs (or atoms) $i$ and $j$ leads to a modified Coulomb potential 
\begin{equation}
\label{eqVgg}
v^{gg}_{ij} = \frac{{\rm erf}(r_{ij} / \sigma_{ij})}{r_{ij}} ,
\end{equation}
where $r_{ij}$ is the interatomic distance, $\sigma_{ij}$ is an effective width, $\sigma_{ij} = \sqrt{\sigma_i^2 + \sigma_j^2}$, obtained from $\sigma_i$ 
and $\sigma_j$, the gaussian widths of atoms $i$ and $j$, respectively. 
Since the polarizability relates the response of a dipole moment to an applied electric field,
the $\sigma_i$ and $\sigma_j$ parameters correspond to the distribution of the dipole moment, and not of the charge.
The width of the gaussian distribution is directly related to the polarizability 
in classical electrostatics~\cite{Mayer-PRB}.

Using Eq.~(\ref{eqVgg}), the dipole interaction tensor for atoms $i$ and $j$ becomes
\begin{eqnarray}
\label{eqTgg}
T_{ij}^{ab} &=& \nabla_{{\bf r_i}} \otimes \nabla_{{\bf r_j}} v^{gg}_{ij} \nonumber \\
&=& -\frac{3 r_{a} r_{b} - r_{ij}^2 \delta_{ab}}{r_{ij}^5}  \left( {\rm erf}(\frac{r_{ij}}{\sigma_{ij}}) - \frac{2}{\sqrt{\pi}}  \frac{r_{ij}}{\sigma_{ij}} e^{-(\frac{r_{ij}}{\sigma_{ij}})^2} \right) \nonumber  \\
&+& \frac{4}{\sqrt{\pi}} \frac{1}{\sigma_{ij}^3} \frac{r_{a} r_{b}}{r_{ij}^2} e^{-(\frac{r_{ij}}{\sigma_{ij}})^2} , 
\end{eqnarray}
where $a$ and $b$ specify general Cartesian coordinates ($x$,$y$,$z$), $r_a$ and $r_b$ are the respective components of the interatomic distance $r_{ij}$,
and $\delta_{ij}$ is the Kronecker delta function.
It can be clearly seen that the above expression reduces the interaction between
dipoles at short distance, in comparison to the bare dipole interaction potential.
Even in the zero-distance limit, it converges to a finite value. 
Therefore, the description of the polarizabilities by a dipole distribution instead
of a point naturally introduces short-range damping effects, which have been so far 
included using \textit{ad hoc} models in the DFT-D and DFT+vdW approaches.

Both DFT-D and DFT+vdW methods use a distance-dependent damping function $f_{\rm{damp}}(r_{ij})$, which
multiplies the $C_6^{ij} R_{ij}^{-6}$ dispersion energy. The function $f_{\rm{damp}}(r_{ij})$ converges
to zero or a small finite value at zero distance between two atoms~\cite{Koide}. At large distances, $f_{\rm{damp}}(r_{ij})$
saturates to unity, typically for distances 20\% larger than the sum of the van der Waals radii of the two atoms.
Besides these two constraints the functional form of the damping is essentially arbitrary. 
Some evidence suggests~\cite{Grimme-damping} that the binding energies are not significantly affected
by the functional form of the damping, as long as two adjustable parameters are used. 
One of these parameters controls the steepness, while the other determines
the onset of the damping in terms of the distance. Typically, only one of these parameters is adjusted
for a given DFT functional by minimizing the error with respect to high-level quantum-chemical binding energies.
Another disadvantage of the damping function, as used in DFT-D and DFT+vdW methods, is the need
to define van der Waals radii, which are not quantum mechanical observables.  
 
Inspection of Eq.~(\ref{eqTgg}) shows that, when using a QHO approximation for the spatial spread of the
polarizability, the damping function is more complicated than a purely multiplicative function.
This complication arises due to the last exponentially decaying term in $T_{ij}^{ab}$. 
Although this conclusion is based on a QHO model, the same conclusion holds for
other models, such as hydrogenic atoms. We conclude that our findings are likely to be valid in general,
meaning that the damping function must be derived from a model Coulomb potential
that naturally accounts for short-range dipole distribution overlap effects~\cite{MBD}. The coupling of
the dispersion energy to a given DFT functional might still require
empirical parameter(s), as we illustrate below. 
However, a seamless coupling may be achieved by
using the range-separation of the Coulomb potential in the calculation of the 
DFT correlation energy. The approach presented in this section can be
employed in the development of such a range-separation procedure.

\section{Efficient evaluation of the ACFD-RPA energy for a system of Quantum Harmonic Oscillators}
The ACFD-RPA approach to the correlation energy has proven to be promising for            
a wide variety of molecules and extended systems~\cite{Fuchs/Gonze:2002,Furche-vanVoorhis,
Furche-RPA,Scuseria/Henderson/Sorensen:2008,Janesko/Henderson/Scuseria:2009,Toulouse/etal:2009,Harl/Kresse:2008,
Harl/Kresse:2009,Lu/Li/Rocca/Galli:2009,Dobson/Wang:1999,Rohlfing/Bredow:2008,Ren/Rinke/Scheffler:2009,Schimka/etal:2010,
Galli-JPCA-2010,Ren/etal:2011}. The largest drawback of ACFD-RPA calculations is their relatively
high computational cost, resulting in a steep increase in the required computational time with system size.  
The conventional scaling of the ACFD-RPA calculations is $N^5$, where $N$ is the number of basis functions.
This can be reduced to $N^4$, when resolution-of-the-identity, or density-fitting, techniques are employed to compute
the four-centered two-electron Coulomb integrals~\cite{Xinguo-beyond-DFT}.

If we only aim at computing the long-range vdW energy, it is possible to associate a single quantum harmonic oscillator (QHO) to
every atom. For such a system, in which the QHOs interact within the dipole approximation, one can circumvent evaluation of the four-centered two-electron 
Coulomb integrals (and costly summations over the virtual/unoccupied functions) via application of the QHO selection rules~\cite{Bade,Donchev,Cole-CFDM}. 
In this section, we will demonstrate that the ACFD-RPA correlation energy for a system of QHOs interacting through a dipole potential is equivalent to the interaction 
energy obtained from diagonalizing the corresponding Hamiltonian matrix. Within this formalism, the computational scaling will be reduced to $N^3$ 
(via the diagonalization step), where $N$ is simply the number of atoms in the molecular system of interest. Further computational savings can be obtained if 
one computes the interaction energy with efficient path integral techniques~\cite{Cao-Berne,Maggs}, instead of diagonalizing the coupled QHO Hamiltonian matrix, 
allowing for the efficient computation of the many-body dispersion energy during Monte Carlo (MC) and molecular dynamics (MD) simulations. 

Throughout the remainder of this Section, we will restrict the derivation to a system of one-dimensional QHOs in order to simplify the notation. The extension to 
three-dimensional QHOs is straightforward and does not alter the conclusions. 


\subsection{The Hamiltonian for a System of Coupled QHOs}

For a system of $N$ QHOs interacting within the dipole approximation, the Hamiltonian can be written as~\cite{MBD}:
\begin{equation}
\hat{H}=-\sum_p^N \frac{\bigtriangledown^2_{{\bf \xi}_p}}{2} +\sum_p^N \frac{\omega_p^2 \mathbf{\xi}_p^2}{2}+
\sum_{p>q}^N \omega_p \omega_q \sqrt{\alpha_p \alpha_q} \mathbf{\xi}_p T_{pq} \mathbf{\xi}_q \, ,
\end{equation}
where $\mathbf{\xi}_p$ represents the  displacement of the $p$-th oscillator from its equilibrium distance weighted by the square root of its mass. 
The first two terms in this Hamiltonian represent 
the kinetic energy and the confining potential corresponding to a set of independent QHOs with unit charge and characteristic frequency $\omega_p$ (\textit{i.e.}, 
the eigenvalues of the non-interacting, single-particle QHO Hamiltonian matrix). The third term describes the interoscillator coupling via the dipole-dipole 
interaction, which also depends on $\omega_p$, as well as $\alpha_p$, the QHO static dipole polarizability, and $T_{pq}$, the $N \times N$ dipole-dipole 
interaction tensor (as defined in Section II). Due to the quadratic (bilinear) dependence of the Hamiltonian on the $\mathbf{\xi}_p$ coordinates, it is 
possible to obtain an exact solution for the QHO interaction energy via diagonalization of the following $C^{\rm QHO}$ matrix:
\begin{equation}
\label{CcQHO}
C^{\rm QHO}_{pq}=\delta_{pq}\omega_p^2+(1-\delta_{pq})\omega_p\omega_q\sqrt{\alpha_p\alpha_q} \, T_{pq} \, .
\end{equation}
The resulting interaction energy, $E_{c,\rm QHO}$, is then computed as the difference between the eigenvalues of the \textit{coupled} system 
of QHOs (obtained via diagonalization of the $C^{\rm QHO}$ matrix) and the eigenvalues of the \textit{uncoupled} system of QHOs (the characteristic 
frequencies), \textit{i.e.}, 
\begin{equation}
\label{EcQHO}
E_{c,\rm QHO}=\frac{1}{2} \sum_i^N ( \sqrt{\bar{\omega}_i^2} - \omega_i) \, .
\end{equation}
From Eq.~(\ref{EcQHO}), the interaction energy for a system of coupled QHOs is characterized by a set of normal modes, whose corresponding frequencies 
are shifted with respect to the non-interacting characteristic frequencies due to the presence of dipole--dipole coupling:
\begin{equation}
\label{CcQHOeval}
\bar{\omega}_i^2 = \omega_i^2 + \Delta_i \, .
\end{equation}

\subsection{The ACFD-RPA Correlation Energy for a System of Coupled QHOs}

After performing the integration over the coupling constant $\lambda$ in Eq.~(\ref{eqRPAl}), the ACFD-RPA correlation energy can be written in the 
following form (an alternative yet equivalent expression to Eq.~(\ref{eqRPA}))~\cite{XinguoJMS2012}:
\begin{equation}
\label{eqRPA2}
E_{c,\rm RPA}=\frac{1}{2\pi} \int_{0}^{\infty} d\omega \, {\rm Tr}[{\rm ln}(1-\chi_0v)+\chi_0v].
\end{equation}
As discussed in Section II, $\chi_0v$ corresponding to a set of coupled QHOs can be written in matrix form as $AT$ (See Eq.~(\ref{atmatrix})) by 
utilizing the QHO selection rules and the inherent locality of the QHO polarizabilities. Since ${\rm Tr}[AT]=0$, the ACFD-RPA correlation energy for 
a system of coupled QHOs can be written as
\begin{equation}
\label{ecrpa}
E_{c,\rm RPA}=\frac{1}{2\pi} \int_0^{\infty} d\omega \, \ln det[C^{\rm RPA}(\ii\omega)] \, ,
\end{equation}
in which the $C^{\rm RPA}$ matrix is defined as follows:
\begin{equation}
\label{CcRPA}
C^{\rm RPA}_{pq}(\ii\omega)=\delta_{pq}+(1-\delta_{pq})\alpha_p(\ii\omega)T_{pq} \, .
\end{equation}

\subsection{Deconvolution of the $\omega_p$ and $\bar{\omega}_p$ Dependencies in the $C^{\rm RPA}$ Matrix}

In order to disentangle the $\omega_i$ and $\bar{\omega}_i$ dependencies in the $C^{\rm RPA}$ matrix, we seek to rewrite $C^{\rm RPA}$ as the product of two
diagonal $\omega$-dependent matrices which separately contain $\omega_p$ and $\bar{\omega}_p$, respectively. This is accomplished by first extracting the free 
QHO polarizabilities in the $C^{\rm RPA}$ matrix via
\begin{equation}
C^{\rm RPA}(\ii\omega)=-A(\ii\omega)B(\ii\omega) \, ,
\end{equation}
where $A_{pq}(\ii\omega)=-\delta_{pq}\alpha_{q}(\ii\omega)$ (\textit{c.f.} Section II) is a diagonal $N \times N$ matrix which only depends on the uncoupled characteristic 
frequencies, $\omega_p$, and 
\begin{equation}
B_{pq}(\ii\omega)=\delta_{pq}\alpha_p(\ii\omega)^{-1} + (1-\delta_{pq})T_{pq} \,
\end{equation}
is a non-diagonal $N \times N$ matrix which depends on both $\omega_p$ and $\bar{\omega}_p$. Since the dipole--dipole interaction tensor is frequency 
independent, one can follow the procedure suggested in Ref.~\cite{applequist} and recast the $B(\ii\omega)$ matrix as
\begin{equation}
B(\ii\omega)=B(0)+D\omega^2 \equiv B(0)+\delta_{pq} (\alpha_p(0)\omega_p^2)^{-1}\omega^2 \qquad
\end{equation}
in which the $D$ matrix has been defined explicitly. In this form, $B(\ii\omega)$ can now be written in terms of a diagonal matrix by solving the 
following generalized eigenvalue problem:
\begin{equation}
B(0) \mathbf{t}_n = \tilde{\omega}_n^2 D \mathbf{t}_n \, ,
\end{equation}
where $\mathbf{t}_n \,\,\,\, (n=1,\ldots,N)$ is a complete set of $N$-dimensional vectors which diagonalize the $D$ matrix, \textit{i.e.}, 
$\mathbf{t}^T_m D \mathbf{t}_n=\delta_{mn}$. This is easily seen if one defines $\tilde{T}=[\mathbf{t}_1,\mathbf{t}_2,\ldots,\mathbf{t}_N]$, 
from which
\begin{equation}
\tilde{T}^T B(\ii\omega) \tilde{T} = \Omega(\ii\omega) \equiv \delta_{pq} (\omega^2+\tilde{\omega}_p^2) \, .
\end{equation}
The $C^{\rm RPA}$ matrix can now be written in terms of the two diagonal matrices $A(\ii\omega)$ and $\Omega(\ii\omega)$:
\begin{equation}
\label{CcRPAAW}
C_{\rm RPA}(\ii\omega)=-A(\ii\omega) (\tilde{T}^{T})^{-1} \Omega(\ii\omega) \tilde{T}^{-1} \, .
\end{equation}
The matrix $A(\ii\omega)$, containing the free polarizabilities, will only depend on the uncoupled QHO frequencies $\omega_p$. The coupled QHO 
frequencies, $\bar{\omega}_p$, on the other hand, will be present inside the $\Omega(\ii\omega)$ matrix through its dependence on the 
$\tilde{\omega}_p$. The relationship between the $\tilde{\omega}_p$ and $\bar{\omega}_p$ can be determined by defining a diagonal 
matrix $F$ such that $F^TF=D$,
\begin{equation}
F_{pq}=\delta_{pq}(\alpha_p(0)\omega_p^2)^{-1/2}
\end{equation}
and observing the fact that $B(0)=F^TC^{\rm QHO}F$, which allows for diagonalization of the coupled $C^{\rm QHO}$ matrix, \textit{i.e.},
\begin{equation}
\tilde{T}^T B(0) \tilde{T} = \tilde{T}^T F^T C^{\rm QHO} F\tilde{T} = \Omega(0) \, .
\end{equation}
Hence, $\Omega(0)$ is the matrix of the coupled eigenvalues of $C^{\rm QHO}$ and $\tilde{\omega}^2_i$ = $\bar{\omega}^2_i$.

\subsection{Frequency Integration and the Contributions from the $\omega_p$ and $\bar{\omega}_p$ Poles}

Before proceeding to the integration over frequency in Eq.~(\ref{ecrpa}), we first consider the logarithm operating on $C^{\rm RPA}$:
\begin{equation}
\ln det[C^{\rm RPA}]=\ln det[-A\Omega]+\ln det[(\tilde{T}^{T})^{-1}\tilde{T}^{-1}] \, ,
\end{equation}
in which Eq.~(\ref{CcRPAAW}) was used. Since $(\tilde{T}\tilde{T}^T)^{-1}=D$ and
\begin{equation}
\label{aomega}
(-A(\ii\omega)\Omega(\ii\omega))_{pq}=\delta_{pq} \frac{\alpha_p(0)}{1+\omega^2/\omega_p^2}(\bar{\omega}_p^2+\omega^2) \, ,
\end{equation}
all that remains in the expression for $E_{c,\rm RPA}$ is $N$ integrals of the form
\begin{equation}
E_{c,\rm RPA}=\frac{1}{2\pi} \sum_{p=1}^N \int_0^{\infty} d\omega \, \ln\left( \frac{\bar{\omega}_p^2+\omega^2}{\omega_p^2+\omega^2}\right) \, ,
\end{equation}
which, after integration by parts and the use of Eq.~(\ref{CcQHOeval}), becomes
\begin{equation}
E_{c,\rm RPA}=\frac{1}{2\pi} \sum_{p=1}^N \int_0^{\infty} d\omega \frac{2\Delta_p \omega^2}{(\omega_p^2+\omega^2) (\bar{\omega}_p^2+\omega^2)} \, .
\end{equation}
This integrand shows both a pair of \textit{coupled} ($\omega = \pm i \bar{\omega}_p$) and \textit{uncoupled} ($\omega = \pm i \omega_p$) QHO poles.
Extending the integral to $-\infty$ by symmetry and closing the integration path in the upper imaginary half plane results in the fact that only the 
QHO poles possessing a positive imaginary component provide a non-zero contribution to $E_{c,\rm RPA}$. By explicitly performing the frequency integration, 
the coupled and uncoupled poles will contribute with a $|\bar{\omega}_i|/2$ and $-\omega_i/2$ term, respectively. From Eq.~(\ref{EcQHO}), the sum of these 
contributions for each of the $N$ QHOs yields $E_{c,\rm QHO}$. 

Hence, the ACFD-RPA correlation energy for a set of QHOs coupled through a dipole--dipole potential is equivalent to the interaction energy that 
one obtains upon exact diagonalization of the Hamiltonian for a system of coupled QHOs. The coupled QHO normal modes naturally include many-body 
effects in complete analogy to the ACFDT-RPA energy. Although Eq.~(\ref{EcQHO}) gives no further insight concerning the nature of these many-body 
effects, we have demonstrated how the coupled QHO model naturally provides beyond-pairwise many-body energy contributions (up to $N$-th order) 
and the RPA response screening (up to infinite order). As a result, diagonalization of the coupled QHO Hamiltonian allows for an effective RPA treatment 
of long-range vdW interactions at a significantly reduced computational cost. As such, the coupled QHO model represents a highly efficient and tractable 
method for the calculation of the many-body vdW energy in large scale systems. Furthermore, we stress that the present results do not depend on the 
choice of the $T$ matrix. Any shape of the dipole--dipole interaction (as long as it remains frequency-independent) would not alter the validity of these 
conclusions.

\section{The importance of many-body effects for the interatomic dispersion energy}

The pairwise dispersion energy is an approximate form that we derived starting from 
the exact ACFD formula for the correlation energy in Eq.~(\ref{eqACFD}). This derivation was based
on two approximations: (i) Analytic integration over the adiabatic connection parameter 
$\lambda$ using $f_{xc}=0$ (RPA) (ii) Second-order truncation of the logarithmic series expansion resulting
from (i). 
While the first approximation was proven to hold exactly for a system of coupled QHOs, the
logarithmic series truncation limits the validity of the pairwise approximation to second-order
perturbation theory.
In this section, we assess the effect of going beyond the pairwise approximation.
This analysis is complementary to our recent work showing the importance of many-body
dispersion energy for a variety of molecules and solids~\cite{MBD,PNAS-2012}. 
The difference here is that our analysis is now based on the rigorous ACFD-RPA expression.

We model each atom $i$ as a QHO as explained in the previous section. The input
parameters $\alpha_0$ and $\omega_0$ in Eq.~(\ref{eqQHO}) are obtained from
first-principles by using the Tkatchenko-Scheffler (TS) method~\cite{TS-vdw}. 
In practice, for a given molecule, we carry out a DFT calculation 
using the PBE exchange-correlation functional~\cite{PBE}. The
resulting self-consistent electron density is then partitioned into
atomic contributions using the Hirshfeld approach~\cite{Hirshfeld}.   
Both $\alpha_0$ and $\omega_0$ are then defined as functionals of 
the atomic electron density. We note in passing that using densities from different functionals
leads to essentially the same final results (for further details see Ref.~\cite{TS-vdw}).
The width $\sigma_i$ of every QHO in Eq.~(\ref{eqQHO}) is determined 
by using Eq.~(\ref{eqTgg}) in the limit of zero-distance
between two dipoles, $\sigma_i=\left(\sqrt{2/\pi} \alpha_i / 3 \right)^{\frac{1}{3}}$~\cite{Mayer-PRB}.

For small and medium size molecules, the TS method yields intermolecular $C_6$ coefficients in excellent
agreement with experimental values obtained from dipole-oscillator
strength distributions~\cite{TS-vdw}. In this case, the input TS polarizabilities $\alpha^{\rm{TS}}$
correspond more closely to the fully interacting response function $\chi_1$ 
than to the non-interacting response function $\chi_0$ used in Eq.~(\ref{eqRPA}). 
Expanding ACFD-RPA formula in terms of the interacting response function $\chi_1$, 
instead of $\chi_0$, leads to the following expression for the correlation energy~\cite{Galli-JCP-RPA-expansion}
\begin{equation}
\label{eqRPAscr}
E_c = - \frac{1}{2\pi} \int_{0}^{\infty}d\omega \sum_{n=2}^{\infty} (-1)^{n} \left(1-\frac{1}{n}\right) {\rm Tr}[(\chi_1 v)^n] .
\end{equation} 

Comparing Eq.~(\ref{eqRPAscr}) to Eq.~(\ref{eqRPA}), it is clear that the second-order
term remains the same, albeit operating now on $\chi_1$ instead of $\chi_0$. This fact 
demonstrates that the origin of the polarizabilities does not modify
the pairwise additive expression for the dispersion energy, further
strengthening our derivation of the pairwise interatomic methods from the ACFD
formula in Section II. In contrast, the coefficients
of the higher-order terms differ significantly between Eq.~(\ref{eqRPAscr}) and Eq.~(\ref{eqRPA}).
In fact, the terms with even powers of $n$ carry a negative sign in Eq.~(\ref{eqRPAscr}), while
all the terms in Eq.~(\ref{eqRPA}) are positive. We note that the modified Coulomb potential
$W^{\prime}$ proposed in Ref.~\cite{MBD} leads to a similar switch in the sign of the many-body energy
contributions.

\begin{table}
\caption{
Performance of different functionals with dispersion energy on the S22 database
of intermolecular interactions. The errors are measured with respect to basis-set
extrapolated CCSD(T) calculations of Takatani \textit{et al.}~\cite{Sherrill-S22}.
Mean absolute relative errors (MARE in \%), standard deviation (SD in kcal/mol), and mean
absolute errors (MAE in kcal/mol) are reported. The postfix ``-2D'' means that the dispersion
energy is added using the second-order expansion of Eq.~(\ref{eqRPAscr}), while
``-$\infty$D'' means that the dispersion energy is computed to infinite order.
The DFT calculations have been carried out with the FHI-aims code~\cite{blum2009}
using a large numeric tier 3 basis set. For DFT, this basis set is converged to
better than 0.05 kcal/mol compared to the basis set limit~\cite{jctc-2011}.
}
\label{tabS22}
\begin{ruledtabular}
\begin{tabular}{lcccccccc}
\hline
\mbox{} & \mbox{$\beta$} & \mbox{MARE} & \mbox{SD} & \mbox{MAE} \\
\hline
PBE-2D          & 2.06 & 11.6\% & 0.84 & 0.64 \\
PBE-$\infty$D   & 2.50 & 7.1\%  & 0.57 & 0.43 \\
PBE0-2D         & 2.12 & 11.4\% & 0.96 & 0.72 \\
PBE0-$\infty$D  & 2.52 & 7.2\%  & 0.66 & 0.52 \\
\hline
\end{tabular}
\end{ruledtabular}
\end{table}

\begin{table}
\caption{
Performance of PBE0 functional including effective pairwise dispersion (PBE0-2D) and the
full many-body dispersion (PBE0-$\infty$D) on the dispersion-bound complexes contained in the S22 database, with respect to the
basis-set extrapolated CCSD(T) calculations of Takatani \textit{et al.}~\cite{Sherrill-S22}.
All values are reported in kcal/mol.
The DFT calculations have been carried out with the FHI-aims code~\cite{blum2009}
using a large numeric tier 3 basis set. For DFT, this basis set is converged to
better than 0.05 kcal/mol compared to the basis set limit~\cite{jctc-2011}.
}
\label{tabDisp}
\begin{ruledtabular}
\begin{tabular}{lcccccccc}
\hline
\mbox{} & \mbox{PBE0-2D} & \mbox{PBE0-$\infty$D} & \mbox{CCSD(T)} \\
\hline
Methane dimer                     & -0.63 & -0.58 & -0.53 \\
Ethene dimer                      & -1.64 & -1.37 & -1.48 \\
Benzene--Methane                  & -1.45 & -1.48 & -1.45 \\
Benzene dimer ($C_{2h}$)          & -1.80 & -2.50 & -2.62 \\
Pyrazine dimer                    & -3.00 & -3.23 & -4.20 \\
Uracil dimer                      & -8.80 & -9.57 & -9.74 \\
Indole--Benzene Stack             & -3.15 & -4.54 & -4.59 \\
Adenine--Thymine Stack            & -9.86 & -11.80 & -11.66 \\
\hline
\end{tabular}
\end{ruledtabular}
\end{table}


Widely used (semi)-local and hybrid functionals in DFT, such as PBE~\cite{PBE}, PBE0~\cite{PBE0-1,PBE0-2},
and B3LYP~\cite{BeckeExchange,LYP}, are relatively successful for the short-range correlation energy. In contrast, our approach based on
the QHO model, is constructed to accurately describe the long-range correlation energy.
A seamless connection between a given DFT functional and the QHO model requires
an explicit modification of the DFT functional correlation hole.
This offers an interesting direction for future work. Here, instead
we introduce an empirical parameter $\beta$ that multiplies the QHO--QHO interaction parameter
$\sigma$ in Eq.~(\ref{eqVgg}). A value of $\beta$ larger than unity corresponds to an
interaction that is shifted to larger distances, effectively capturing only the
long-range part of the correlation energy.

In order to assess the accuracy of different approximations to the ACFD formula, we
have chosen to use the S22 database of intermolecular interactions~\cite{S22}, a widely used 
benchmark database with binding energies calculated by a number of
different groups using high-level quantum chemical methods~\cite{S22,Sherrill-S22}.
In particular, we use the recent basis-set extrapolated CCSD(T) binding energies
calculated by Takatani \textit{et al.}~\cite{Sherrill-S22}.
These binding energies are presumed to have an accuracy of $\approx$
0.1 kcal/mol (1\% relative error). This level of accuracy allows an
unbiased assessment of approximate approaches for treating dispersion interactions.
Table~\ref{tabS22} summarizes the results of our calculations on the S22
database. We used two different non-empirical DFT functionals: PBE~\cite{PBE} and PBE0~\cite{PBE0-1,PBE0-2}. 
For every functional, we have carried out two tests: (i) using the second-order
expansion of Eq.~(\ref{eqRPAscr}), and (ii) using the full (infinite) series
in Eq.~(\ref{eqRPAscr}).
The $\beta$ parameter has been adjusted for every combination of functional
and method. We note that the approach presented here does not require the definition of
van der Waals radii---all the necessary information is contained in the
input frequency-dependent polarizabilities and the adjusted $\beta$ parameter.

An analysis of the performance of different methods on the S22 database in Table~\ref{tabS22} reveals 
that the addition of the pairwise dispersion energy in the QHO approximation yields a mean
absolute error (MAE) with respect to CCSD(T), which is typically 
a factor of 2 larger than for DFT-D and DFT+vdW methods~\cite{Grimme-D3,jctc-2011}. 
We note that the MAE can be reduced significantly by using a steeper functional 
form for the damping of the short-range interaction~\cite{MBD}. The QHO approximation
describes every atom as a single gaussian function, which leads to a smooth
damping of the dispersion energy at short distances. Thus, even hydrogen-bonded
systems are significantly stabilized. Interestingly, the inclusion of
the infinite order dispersion energy beyond the pairwise approximation 
noticeably reduces the errors and increases the $\beta$ value for every
tested functional. Both of these results are desirable, since a larger
value of $\beta$ means that the dispersion energy is shifted
to larger distances. 
In particular, the error is reduced for \emph{all}
dispersion-bound systems when going from PBE0-2D to PBE0-$\infty$D
as shown in Table~\ref{tabDisp} (the exception is the benzene-methane dimer,
where both PBE0-2D to PBE0-$\infty$D yield essentially the same results).
The most pronounced deviation between PBE0-2D and PBE0-$\infty$D
is observed for the largest dispersion-bound adenine--thymine complex.
The CCSD(T) method yields a binding energy of --11.7 kcal/mol, while
PBE0-2D yields --9.9 kcal/mol, and PBE-$\infty$D improves the estimate to --11.8 kcal/mol.
This agrees with our previous findings using an interatomic many-body
dispersion method~\cite{MBD}, and demonstrates that the many-body dispersion energy
becomes significant as the molecule size increases.

\section{Conclusions}
The widely used DFT-D and DFT+vdW methods that compute the dispersion energy
as a pairwise sum over atoms have been derived from the quantum mechanical
adiabatic connection formula. This derivation puts interatomic dispersion
methods on a firm theoretical basis. We have shown that the damping of the 
dispersion energy at short interatomic distances is connected to the 
spatial spread of the dipole moment. 
We have also demonstrated that the non-additive many-body effects beyond 
the pairwise approximation play an important role for the binding energies
of dispersion-bound complexes.
Moreover, given the equivalence between the exact and RPA treatment of the coupled QHO 
model, the full many-body dispersion energy can be efficiently computed with a single 
matrix diagonalization.

There are many avenues remaining for future
work, including, for example,
(i) different approximations to the ACFD formula, 
(ii) the role of input polarizabilities into the ACFD formula, 
(iii) the role of anisotropy and localization in the input polarizabilities,
(iv) the role of higher multipole moments in the response function, and
(v) improving the coupling between DFT and the long-range dispersion energy.
The adiabatic connection formula provides a powerful framework for the development of accurate
and efficient approaches for computing the correlation energy
in general and the dispersion energy in particular.

\begin{acknowledgments}
All authors acknowledge the European Research Council (ERC Starting Grant \texttt{VDW-CMAT}),
and insightful discussions with Xinguo Ren and John F. Dobson.
\end{acknowledgments}

\bibliography{literature,beyondDFT}
\end{document}